\documentclass{ieeeaccess}
\usepackage{cite}
\usepackage{amsmath,amssymb,amsfonts}
\usepackage{algorithmic}
\usepackage[pdftex]{graphicx}
\usepackage{textcomp}
\usepackage{caption,setspace}
\usepackage{ctable}
\usepackage{url}

\def\BibTeX{{\rm B\kern-.05em{\sc i\kern-.025em b}\kern-.08em
T\kern-.1667em\lower.7ex\hbox{E}\kern-.125emX}}

\begin{document}
\history{Date of publication xxxx 00, 0000, date of current version xxxx 00, 0000.}
\doi{XXXXXX}

\title{Towards Trustworthy DeFi Oracles: Past, Present and Future}

\author{
\uppercase{Yinjie Zhao},
\uppercase{Xin Kang} \IEEEmembership{Senior Member, IEEE},
\uppercase{Tieyan Li} \IEEEmembership{Member, IEEE},
\uppercase{Cheng-Kang Chu} \IEEEmembership{Member, IEEE},
\uppercase{Haiguang Wang} \IEEEmembership{Senior Member, IEEE}
}

\address{Digital Identity and Trustworthiness Laboratory, Huawei Singapore, Singapore 138588}

\corresp{Corresponding author: Xin Kang (kang.xin@huawei.com)}

\begin{abstract}

With the rapid development of blockchain technology in recent years, all kinds of blockchain-based applications have emerged. Among them, the decentralized finance (DeFi) is one of the most successful applications, which is regarded as the future of finance. The great success of DeFi relies on the real-world data which is not directly available on the blockchain. Besides, due to the deterministic nature of blockchain, the blockchain cannot directly obtain indeterministic data from the outside world (off-chain). Thus, oracles have appeared as a viable solution to feed off-chain data to blockchain applications. In this paper, we carry out a comprehensive study on oracles, especially on DeFi oracles. We first briefly introduce the application scenarios of DeFi oracles, and then we talk about the past of DeFi oracles by categorizing them into several types based on their design features. After that, we introduce five popular DeFi oracles currently in use (such as Chainlink and Band Protocol), with the focus on their system architecture, data validation process, and their incentive mechanisms. We compare these present DeFi oracles from their data trustworthiness, data source trustworthiness and their overall trust models. Finally, we propose a set of metrics for designing trustworthiness DeFi oracles, and propose a potential trust architecture and a few promising techniques for building trustworthiness oracles.
\end{abstract}

\begin{keywords}
DeFi, Oracles, Blockchain, Trustworthiness
\end{keywords}

\titlepgskip=-15pt

\maketitle

\section{Introduction}

\subsection{Background and Motivation}
Blockchain has appeared in the sight of researchers and the public as a hot topic in the last decade, enabling immutable and decentralized transactions recording.. Blockchain is widely applied in the field of decentralization and digitalization, such as decentralized digital identity, decentralized finance (DeFi), decentralized organization. However, the blockchain system is a closed system by design, which cannot natively receive off-chain data from outside world. This is acceptable for the first generation blockchain, such as bitcoin, due to its simple application scenarios. However, for the second generation blockchain, such as Ethereum, the aforementioned issue limits the application of smart contract, which executes pre-determined algorithms once certain pre-defined conditions are fulfilled. This also limits the development of DeFi which is mainly built on smart contracts.

Thus, oracles appeared as a promising technology to solving this issue, i.e., collecting off-chain information data and providing specific data to on-chain smart contracts. Due to their importance, oracles gradually became the key components of DeFi as data providers and as an interface between the blockchain and the off-chain world. Oracles are involved in many circumstances in DeFi, including decentralized exchange, DeFi lending, synthetic assets (stable coins, etc.), insurance, digital payment (cross-border payment, etc.), fulfilling the demand of price data feeding.

In the past few years, DeFi token market capitalization increases largely, reaching more than 150 billion USD by the time of writing \cite{TT1}. In this situation, oracles are given rising importance due to its contribution to asset price reporting, cryptocurrency pricing and so on. As a result, all kinds of oracles have shown up in the DeFi market. However, the trustworthiness of these oracles have not been discussed and investigated in existing literatures. On the other hand, the trustworthiness of oracles is very important since it directly affects the trustworthiness of the DeFi projects, and it can be regarded as the trust root of smart contracts. Thus, how to guarantee the trustworthiness of oracles is of great importance to the blockchain and DeFi world.

\subsection{Related Works}

There are some existing works providing framework of blockchain oracles. On one hand, oracle can be categorized by data transmission process, such as data sources, data transmission approaches and directions. By data sources, oracles were categorized mainly into software and hardware source ones in \cite{FF2} and \cite{MM1}, as well as human source ones in \cite{LL2} and \cite{LL1}. By data transmission direction, oracles were categorized into inbound and outbound ones by works including \cite{FF2},  \cite{LL2}, \cite{LL1}, \cite{NN1}. Inbound oracles transmit data from off-chain side to on-chain side, and vice versa for outbound oracles. Further, by data transmission approach, oracles can be categorized into pulling and pushing based ones as pointed out by \cite{NN1} and \cite{KK1}. Pulling based oracles are executed at data requester side while pushing oracles are executed on data provider side. \cite{JJ2} categorized oracles into centralized and consensus ones by their degree of centralization of data feeds. \cite{AA1} categorized the oracles into voting-based and reputation-based ones and their respective detailed divisions of data validation mechanisms. \cite{UU1} categorized them into prediction markets, centralized data feeds and oracle networks. \cite{GG2} divided data on-chaining into TLS-based, enclave-based and voting-based types.  \cite{LL1} also proposed witness mechanism or reputation algorithms, security and privacy. \cite{AA1} also pointed out the future research direction of oracles, such as lower cost, higher speed, decentralization, and security. \cite{UU1} suggested that oracles should emphasis on data authenticity, integrity, confidentiality and availability. \cite{BB1} suggested that oracles should have more transparency, accountability and operational robustness. However, for the DeFi industry, data trustworthiness is the most important thing, and all those works focused on the functionality of oracles while paying little attention to the trustworthiness.

\subsection{Contribution and Novelty}

The aforementioned papers gave an overview of categorization of oracles, which mainly focused on the approach of data transmission and data validation themselves. However nowadays, in an era of emphasized data trustworthiness, the discussion and investigation on the trustworthiness of data source and the trustworthiness of data itself is lacked. It is necessary and important to have further up-to-date research to study the data feeding and validation from the trustworthiness perspective. To fill this gap in the field, in this paper we contribute in the following ways:

1) We provide a comprehensive overview on the existing oracles, and categorize these oracles by their data processing and validation process methods and provide visions and analysis of them from a trustworthiness point of view.

2) We provide a detailed analysis of existing popular DeFi oracles, such as Chainlink and Band Protocol. We investigate their design target, system architecture, data feeding scheme, date aggregation scheme, incentive mechanisms, based on which we analyze their trust models and trustworthiness. We also make a detailed comparison on the similarity and difference of their trust models and their trustworthiness.

3) We summarize and propose a set of metrics for designing trustworthy DeFi oracles. We propose a potential trust architecture for DeFi oracles and provide our vision and some viable technical solutions on how to enable the trustworthiness of the data source and the trustworthiness of the data itself.

In Section II, we demonstrate applications of oracles in DeFi with actual DeFi projects. In Section III, we analyze and give a categorization of DeFi oracles, giving analyzation and comparison among them. In Section IV, we give a detail analysis of current popular DeFi oracles and their working mechanisms. We compare different categories of oracles. In Section V, we give our point of view on the future development of trustworthy DeFi oracles.

\section{Application of Oracles in DeFi}

Oracles in DeFi function mainly serve as price data providers. They are widely applied in decentralized exchange, DeFi lending, synthetic assets (such as stable coins), insurance, digital payment (such as cross-border payment), and etc.. In this section, we introduce the application of oracles to DeFi, including in decentralized exchange, synthetic assets and DeFi lending.

\subsection{DeFi Lending}

\begin{figure}
\centerline{\includegraphics[width=20pc]{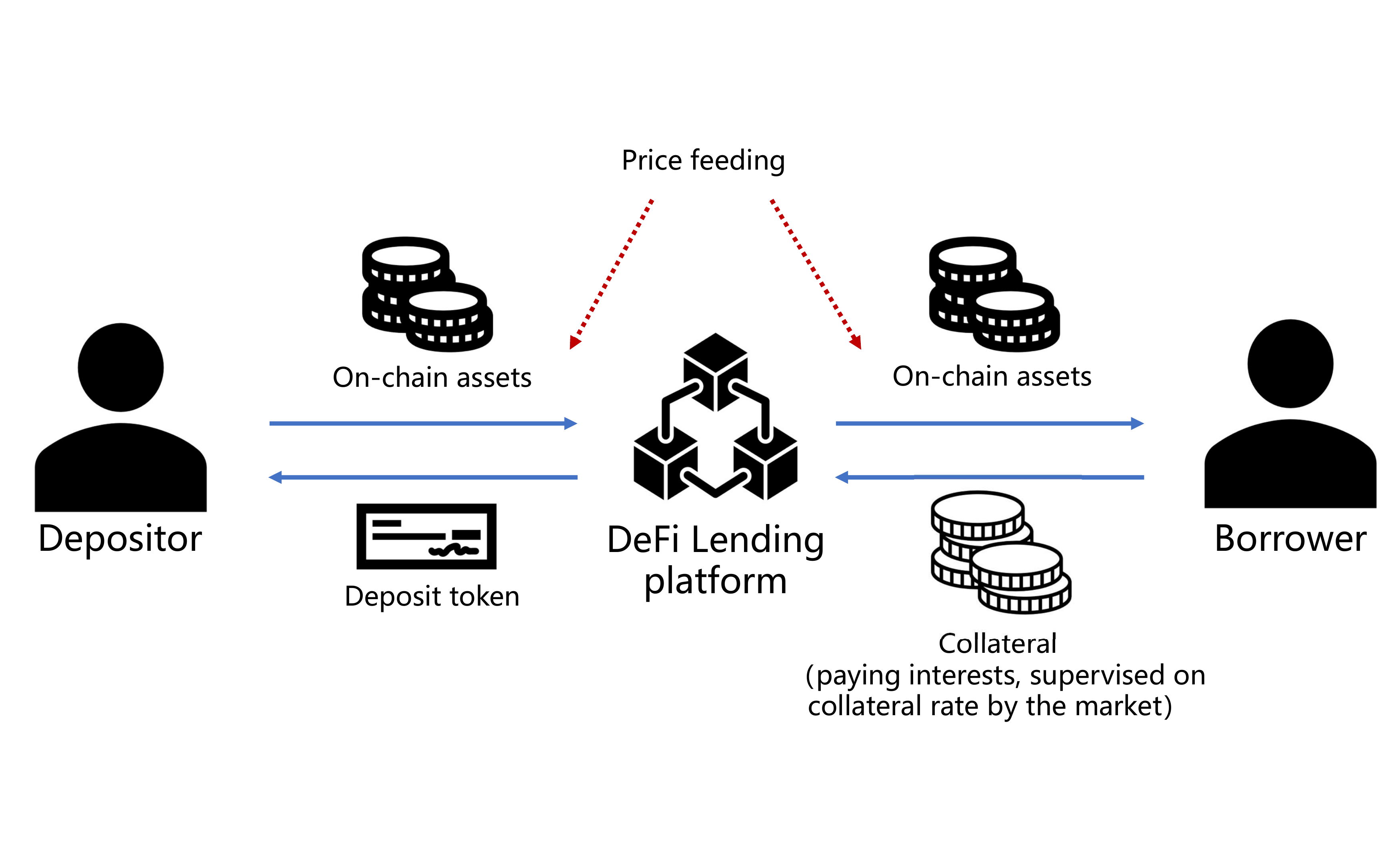}}
\caption{DeFi lending}
\label{defi_lending_label}
\end{figure}

In a DeFi lending platform, price data is fed from oracles in complicated processes (Flowchart shown in Figure \ref{defi_lending_label}). Similar to lending in traditional finance market, borrowers need to collateralize some assets of certain value in case of users not returning the borrowed assets. The collateral rate and amount of deposit token issued are determined by the price data provided as an important input parameter.

\begin{figure}
\centerline{\includegraphics[width=20pc]{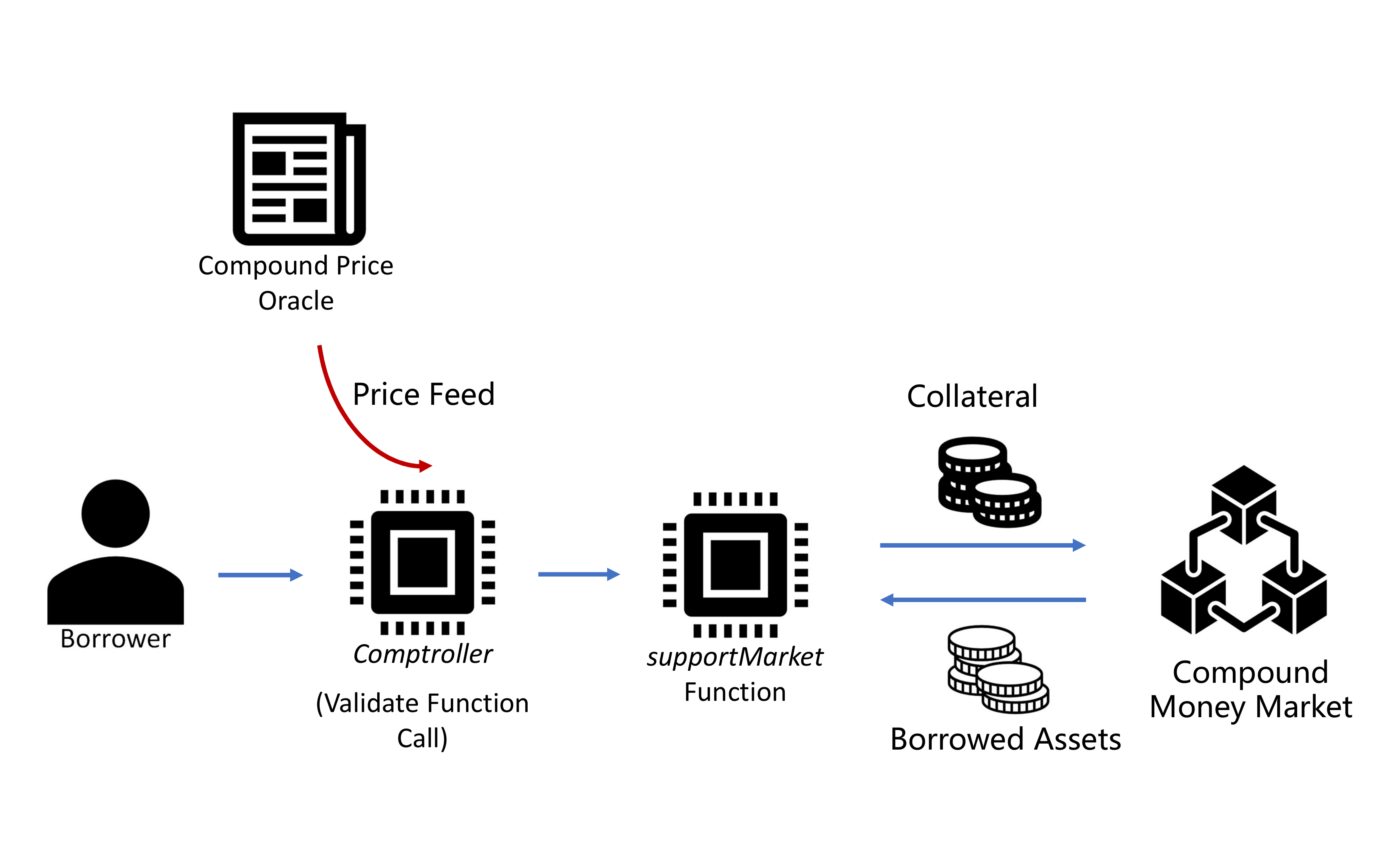}}
\caption{Compound Lending}
\label{compound_label}
\end{figure}

An example of DeFi lending is Compound. In Compound, to borrow an asset, users need to collateralize certain DeFi assets at certain rate against the borrowed assets. Users operate on an asset with a function {\it supportMarket}, which is validated and enabled by a smart contract {\it Comptroller}. Comptroller validates users¡¯ operation via supportMarket by checking collateral rate with price data fed from Compound¡¯s own price oracle, and with valid collateral rate, a lending process is enabled. Oracle, in this circumstance, is the core functionality of the lending platform \cite{II2}, as shown in Figure \ref{compound_label}.

\subsection{Synthetic Assets}

Synthetic assets are created by tracking the value of certain original assets and mint them into new assets by collateralizing, similar to derivatives in traditional finance. Oracles provide price data to calculate the value of DeFi assets and decides the collateral amount on certain assets, so that users can mint new assets by following proper amount of collateral. The collateral rate and amount, as key component of the system operation, is decided by the oracles.

\begin{figure}
\centerline{\includegraphics[width=20pc]{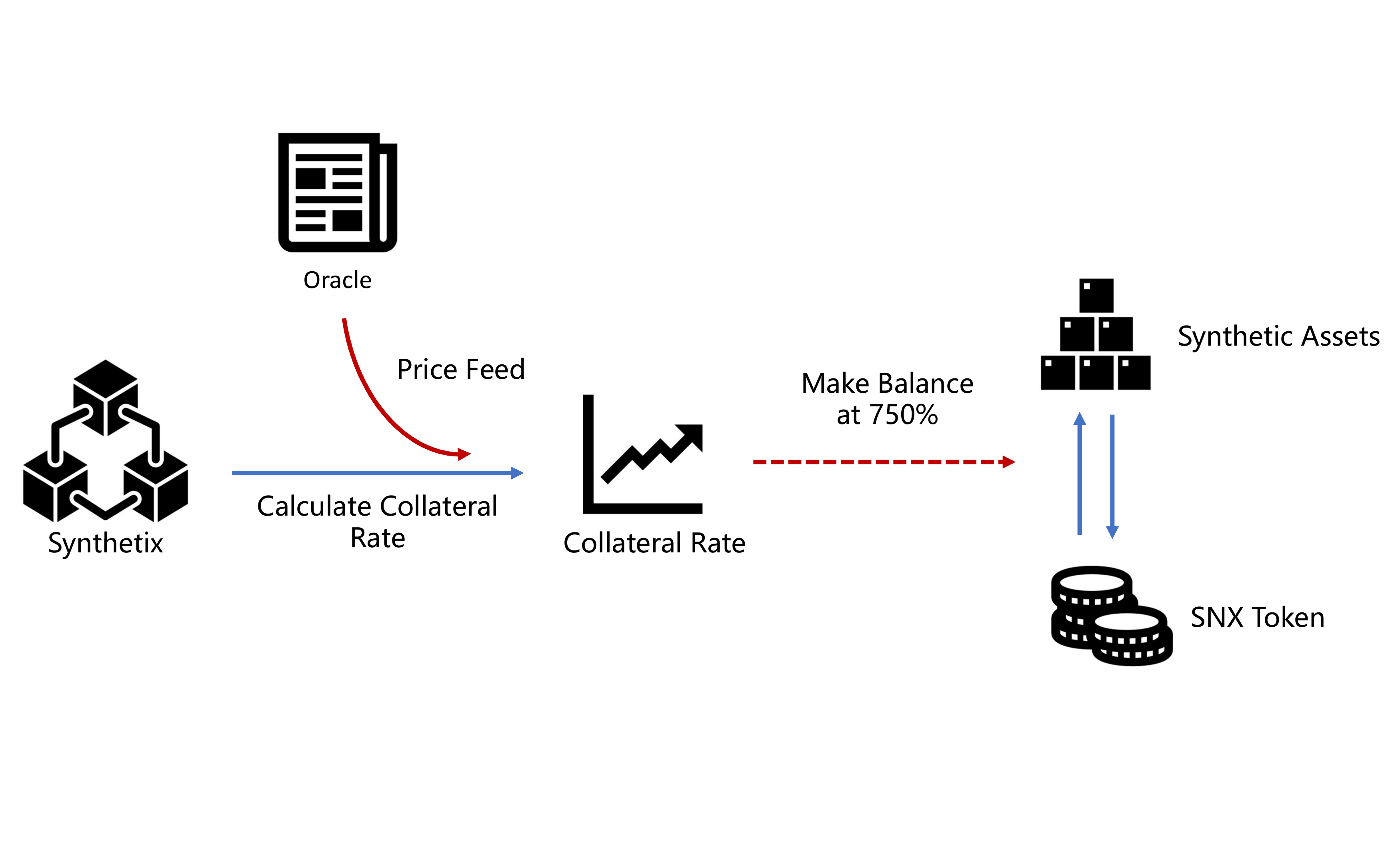}}
\caption{Synthetix Flowchart}
\label{synthetix_label}
\end{figure}

An example is Synthetix, a synthetic asset platform. In Synthetix, users have to collateralize SNX token (the native token of the platform) to Synthetix Exchange to obtain synthetic assets, and a collateral rate of 750\% is required, namely the value of the collaterals should be 7.5 times that of mint assets. Such proportional relationship is supervised by obtaining the price data of assets from oracles\cite{MM2}. For example, if a user mint 10 units of synthetic asset {\it A} worth 100 USD (10 USD per unit of A), then 250 SNX worth 750 USD (3 USD per SNX) need to be collateralized. If now SNX experiences a depreciation, with its price down to 2 USD per SNX, then the value of collateral decreases to 500 USD, lower than the required 750 USD. Oracles feed the price data to the system, which calculates the value change. In this case the user will be required to increase the SNX collateralized or burn out certain amount of asset A to rematch the collateral rate, otherwise not being able to liquify the SNX collateralized. As shown in Figure \ref{synthetix_label}, oracles (mainly Chainlink for Synthetix) play an important part in DeFi synthetic assets.

\subsection{Insurance}

Due to the high volatility of DeFi asset value, users holding on DeFi assets face a high risk of incurring loss. Therefore, insurance projects come to demand. Similar to insurance in traditional finance world, users anticipate risk of high value with relatively small cost.

\begin{figure}
\centerline{\includegraphics[width=20pc]{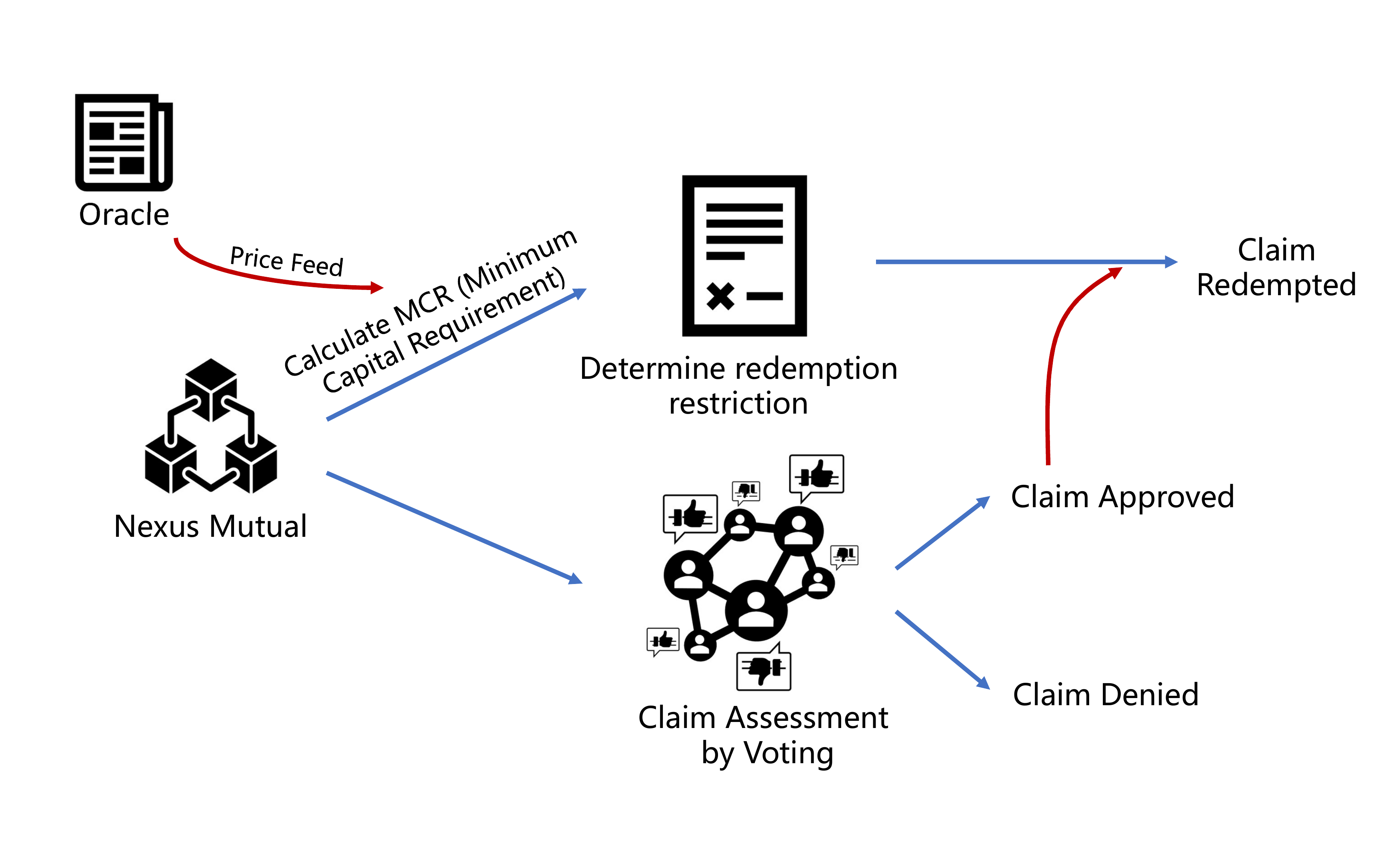}}
\caption{Nexus Mutual Insurance Claiming Flowchart}
\label{nexus_mutual_label}
\end{figure}

An example is Nexus Mutual, a DeFi insurance project. To be covered by the insurance, clients need to collateralize certain amount of assets to reach the Minimum Capital Requirement (MCR), the value and amount of which is determined by price data provided by oracles. In addition, there is a claim assessment process by voting to determine whether to approve the claim of an insurance. If the claim approved, the claim will be executed under the redemption restriction, as shown in Figure \ref{nexus_mutual_label}.

\section{Past of Defi Oracles}
In this section, we compare the difference between centralized and decentralized oracles, and their respective advantages over each other. We categorize decentralized oracles into four categories, based on their data validation process (i.e. source of data validity).

\subsection{Pioneer: Centralized Oracles}
Centralized oracles are usually trusted third party data providers backed by authorities or trustworthy parties (e.g. government authorities, large company with benign reputation, etc.), and usually have single data source. In centralized data providing, data are validated from both software and hardware perspective. The data acquired from local storage of data providers are secured by trusted execution environments (TEEs), which isolate data from the operating system of untrusted local device, and the data transmitted online are secured by data transmission protocols to ensure that the data is not tampered or lost.

\begin{figure}
\centerline{\includegraphics[width=20pc]{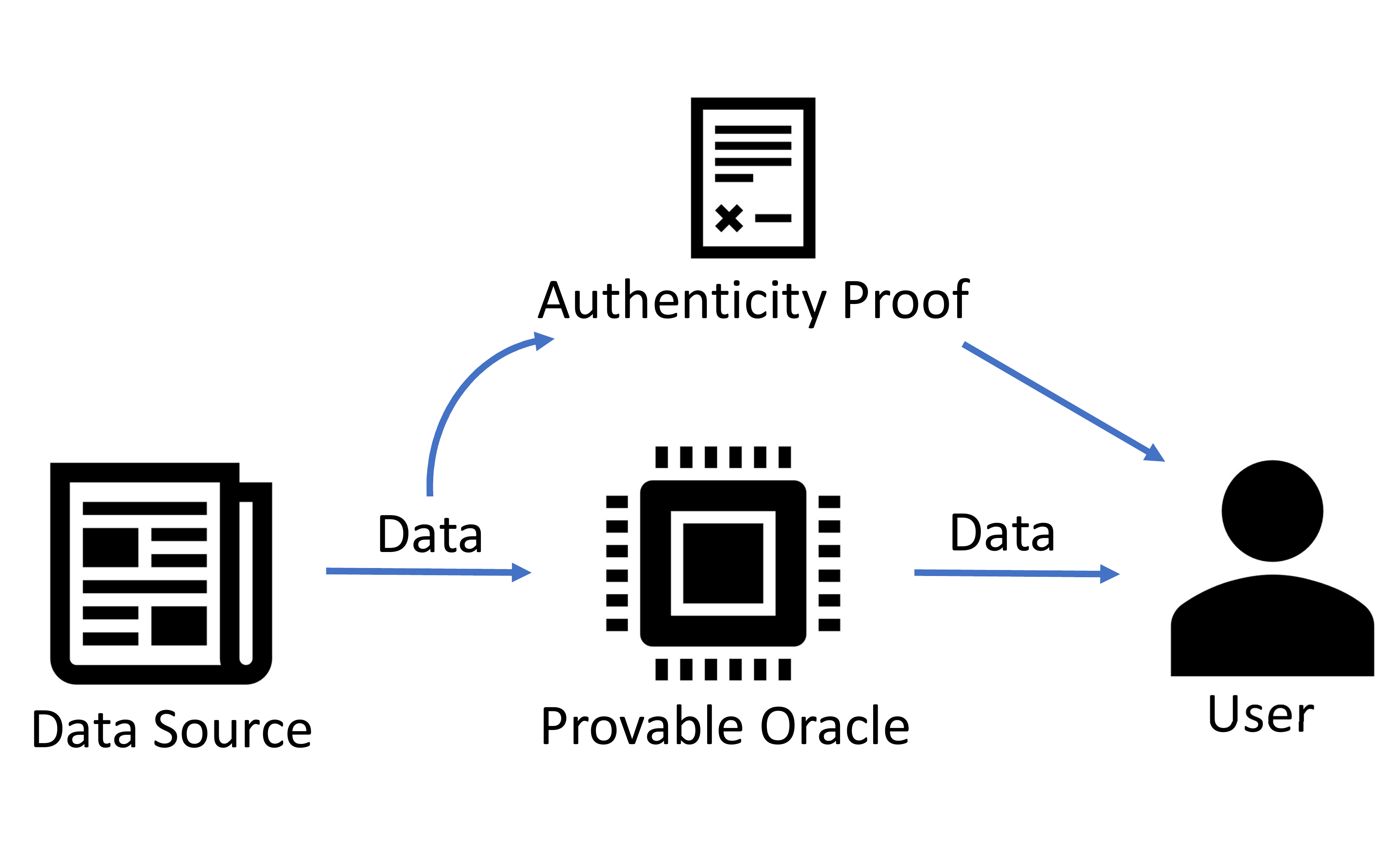}}
\caption{Flowchart of Provable oracle data processing}
\label{provable_label}
\end{figure}

{\it Provable} is an example of centralized oracles, providing users with an authenticity proof of the data along with the data fed. As introduced by the Provable whitepaper, the authenticity proof mainly includes 3 types: the TLSNotary Proof, Android Proof and Ledger Proof. These proofs enable users to audit untampered process of data transmission and are supported by software and HTTP protocols and hardware (namely TEEs) (Flowchart shown in Figure \ref{provable_label}). Provable claims that with authenticity proof users can receive data from Provable without trust. Despite the fact above, users still withstand the probability of Provable¡¯s protocols not reliable or failing at data feeding, for example, encountering system failure, or may be attacked. Such risks are accepted by Provable users and backed by Provable company¡¯s credit\cite{CC2}. Provable now has partnership with IT research companies (Gartner etc.), banking companies (Intesa Sanpaolo, EY etc.), blockchain venture companies (Coinsilium etc.) and so on.

\subsection{Development: from Centralized to Decentralized Oracles}
With the development of DeFi oracles, decentralized oracles began to show their advantage, with increasing application in DeFi. Decentralized oracles obtain data from multiple data sources with certain data validating mechanisms to remove or reduce the impact of malicious data, for example, Chainlink and Nest Protocol.

In comparison, centralized oracles usually have advantage in data processing speed compared to decentralized ones and are easier to construct and realize, while decentralized oracles have better scalabilities and lower failure risks due to the fact of having multiple data sources and decentralized data processing mechanism. Therefore, centralized oracles are more applicable to circumstances with high time-effectiveness requirement than decentralized ones, while decentralized oracles are more applicable to higher adversarial risks situations. Furthermore, decentralized oracles fit better with the ideology and development trend of scalability and decentralization of DeFi. The comparison is summarized in Table \ref{comparison_cenVSde_label}.

\begin{table*}[htbp]
\centering
\caption{Comparison between Centralized and Decentralized Oracles}

\begin{tabular*}{17.1cm}{p{5.2cm}p{5.2cm}p{5.2cm}}

\specialrule{.1em}{.05em}{.05em}
Criteria&
Centralized Oracles&
Decentralized Oracles \\
\specialrule{.1em}{.05em}{.05em}
Data Feeding Mechanism
&Single trusted third party
&Multiple decentralized data sources
\\
\hline
Feasibility
& Relatively higher
& Relatively lower
\\
\hline
Performance
&Higher time-efficiency and data throughput
&Lower time-efficiency and data throughput
\\
\hline
Risks
&Low scalability; single node failure risk
&Strong scalability; resistant to single node failure risk
\\
\hline
Examples
& Provable, etc.
&Chainlink, Band Protocol, NEST Protocol, etc.
\\
\hline
Applicable conditions
&Higher time-efficiency requirement; lower risk-tolerance
&Lower time-efficiency requirement; higher risk-tolerance
\\
\specialrule{.1em}{.05em}{.05em}
\end{tabular*}
\label{comparison_cenVSde_label}
\end{table*}

\subsection{Decentralized Oracles: Data Validation mechanisms}
\begin{figure}
\centerline{\includegraphics[width=20pc]{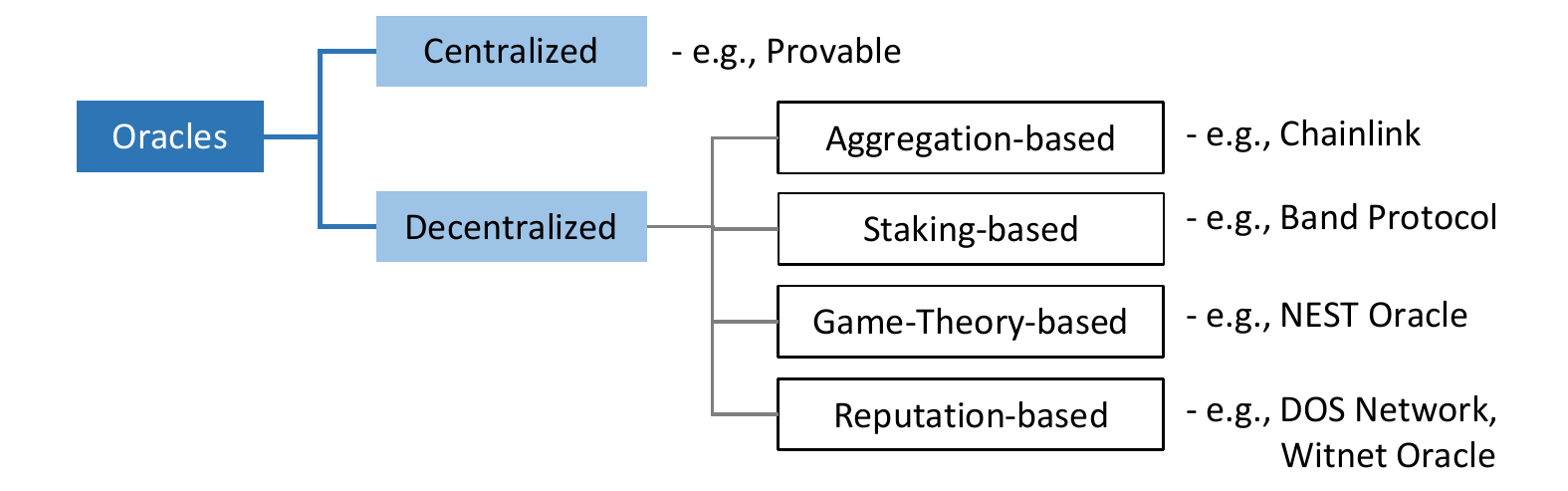}}
\caption{Categorization of DeFi oracles}
\label{categorize_label}
\end{figure}

Decentralized oracles have miscellaneous methods processing and integrating data obtained from providers. In this paper, we categorize the data processing methods into 4 main categories: aggregation-based processing, staking-based processing, game-theory-based processing, and reputation-based processing (as shown in Figure \ref{categorize_label}). Meanwhile, it cannot be ignored that oracles may have a combination of multiple mechanisms mentioned above.

\subsubsection{Aggregation-based processing}
Aggregation-based data processing use predefined logic and mechanisms to obtain a deterministic result from multiple data sources, regardless of individual data quality, cancelling out the impact of malicious data and other adversarial actions by aggregating them with other normal data.

Data aggregation may take miscellaneous methods, including taking the medians, mean value or mode value of the data. One of the commonly used method is taking the median to aggregate the data and is done along with other validation methods including adding an upper bound tolerance of price variance so that the price cannot change in an abnormally large rate (e.g., the Compound), adding time limitation of data validation so that too newly added data or obsolete data is not used for the price reporting (e.g., the AmpleForth) and so on.

\subsubsection{Staking-based processing}
Staking-based data processing produce the trustworthiness of data processing by requiring participants to stake certain amount of assets to enable economic penalty of malicious action to raise the cost of malicious behaviors. For example, Band Protocol has a data validation mechanism of selecting validators with most staking to run the data feeding task, and users conducting adversarial actions will be slashed (meaning his staking reduced by certain amount), and the more stake a validator has, the more likely he will be selected for the data feeding job, and less likely that he behaves dishonestly against his staking. Band Protocol is an example, requiring staking of validators. Another example is the Tellor\cite{BB2}, where miners of a new block has certain stake requirement and will be slashed if the data determined by the block he mined is successfully overthrown by later disputation process.

\subsubsection{Game-theory-based processing}
Game-theory-based data processing provides users and data providers with economic incentives to act in non-adversarial ways. In such systems, users have an overall higher mathematical expectation for benefit from conducting non-adversarial actions than adversarial ones, or a lower expectation for cost of non-adversarial ones than adversarial ones. An example is the NEST Protocol, which constructs a price-chain mechanism. In the system, any user is able to raise a price of assets and claim it as a new block of the price chain. To ensure the uploaded price is not malicious, the new price raiser will provide collateral at the price he raised for a period of time, and if the price is not reasonable for the market within the period, another new-coming user is able to conduct arbitrage at the price and liquify the collaterals and raise a reasonable price while causing loss to the user raising unreasonable price \cite{DD1}.

\subsubsection{Reputation-based processing}
Reputation-based data processing aims at filtering data providers with reputation system, punishing the adversarial nodes and restricting their chance of participation as data providers by reducing their reputation with specific mechanism. {\it DOS Network} (Decentralized Oracle Service Network) is an example of reputation-based data processing. The system provides a service of calculating the {\it QoS score (Quality of Service)} and reducing the score of low-quality nodes, while removing certain nodes with overly abnormal scores\cite{CC1}. Witnet is another example, giving nodes reputation scores, and increment or decrease the score according to the performance of them\cite{QQ1}.

\section{Present of DeFi Oracles}

In this section, we introduce and analyze five active oracles mechanisms, including Chainlink oracle (aggregating-based), Band Protocol oracle (staking-based), NEST Protocol (game-theory-based), DOS Network (reputation-based) and Witnet oracle (reputation-based), and give a brief comparison among them.

\subsection{Chainlink Oracles}
Chainlink oracle is an aggregation-based oracle with the{\it Decentralized Oracle Network} (DON) aggregating data from different data providers and it is one the most popular decentralized oracles in DeFi.

\subsubsection{Design Criteria}
According to the Whitepaper of Chainlink \cite{EE1}, the main goals includes {\it hybrid smart contracts} (functioning as a framework, integrating the computational power both on-chain and off-chain, realizing the full potential of smart contracts), {\it abstracting away complexity} (simplifying user experience), {\it scaling} (meeting the demand of growing scales of the system), {\it confidentiality} (protecting the privacy of users while maintaining the transparency of the system), etc., while the core goal is the first one, which is to provide a hybrid smart contract framework with other goals in service of the first.

\subsubsection{System Architecture}

\paragraph{{\it Decentralized Oracle Network} (DON)}
DON is the core framework of Chainlink as a decentralized oracle. As the name of it indicates, DON is a network of oracles, receiving data from an open network of data providers, processing the data by aggregating to obtain a relatively trustworthy result, and returning them to the users. Furthermore, DON also functions as a key component of transaction execution mechanism (TEF) as mentioned later.

DON mainly consists of executables, adaptors, and storage. Executables are algorithms of DON that are in charge of predetermined algorithms execution including aggregation and has key components including {\it logic} and {\it initiator}. {\it Logic} is deterministic algorithm that executes data aggregation, similar to a smart contract (for example, a {\it logic} L can be set to calculate the median of a given set of data received), while {\it initiator} triggers L under certain determined circumstances. Adaptors define methods and APIs transiting data between outside data providers such as web servers and external storage and DON. Storage of DON stores data in the network, while DON can also have external cloud or decentralized storage \cite{EE1}.

\paragraph{{\it Transaction Execution Framework} (TEF)}
Chainlink not only aims at obtaining and providing trustworthy data but is also proposing its own architecture for on-chain transaction, {\it hybrid smart contracts}, by dividing a smart contract into a {\it Hybrid Contract} consisting of a DON logic off the main chain and an anchor contract on the main chain. The DON logic receives the transactions of users and executes them in the DON and is faster than main-chain transaction execution since it can natively visit oracle data in the DON. The transaction executed will be periodically updated to the main chain. The anchor contract has higher trustworthiness than the anchor contract since it is regarded as immutable on the main chain. Therefore, the anchor point contract is used for asset custodies, syncing verification and as a guard rail of the DON logic.

Furthermore, Chainlink proposed {\it Fair Sequencing Services} (FSS) to cooperate with its TEF, which aims at achieving one of its goals, {\it order-fairness for transactions}. With FSS, the sequence of transactions put on-chain will no longer be determined by economically incentivized users (miners, validators, etc.), but by pre-designed deterministic algorithms so that Miner Extractable Value (MEV) can be reduced.

\begin{figure}
\centerline{\includegraphics[width=20pc]{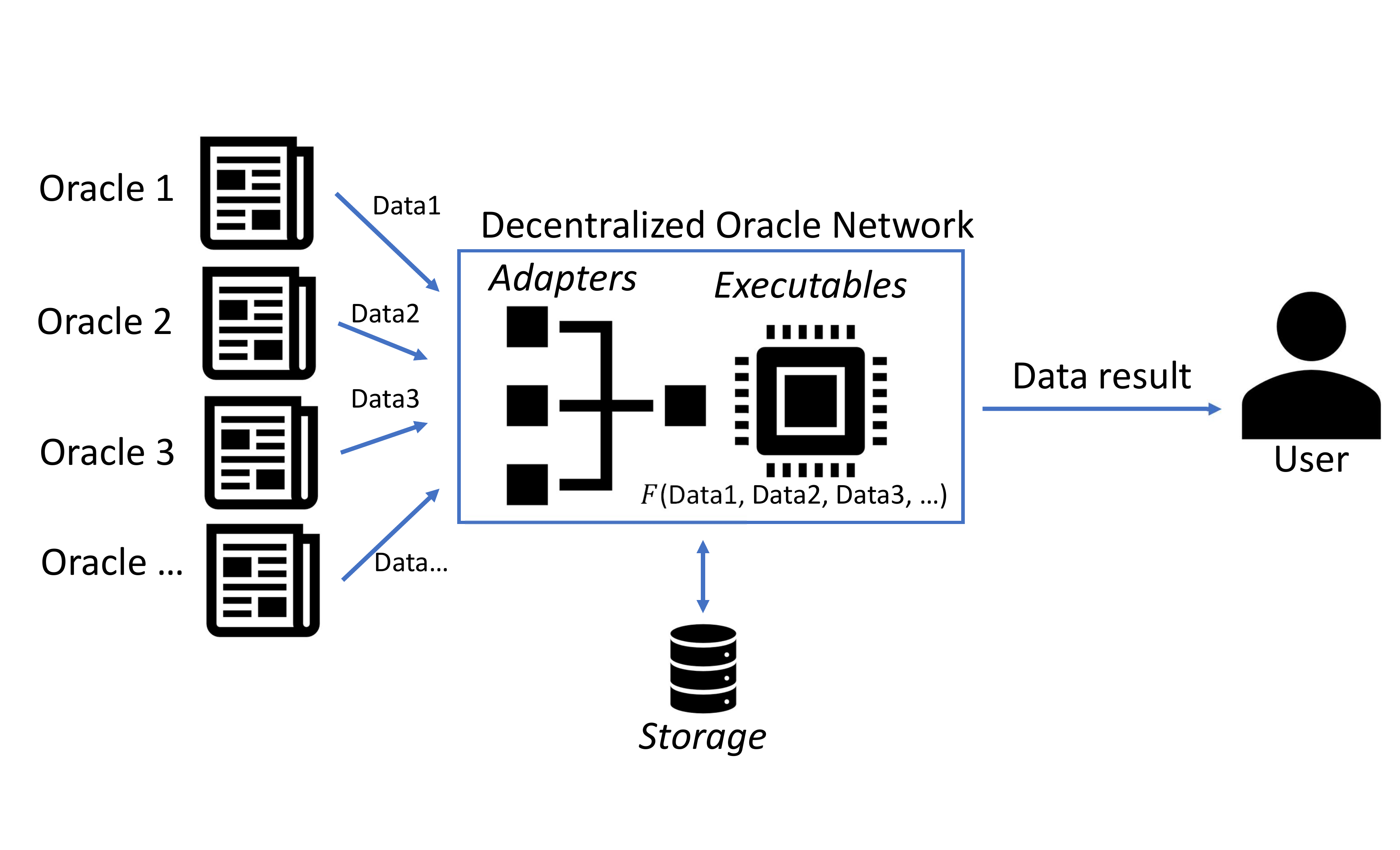}}
\caption{Flowchart of Chainlink oracle data processing}
\label{chainlink_label}
\end{figure}

\subsubsection{Data Validation Process}
In DON, data is mainly validated and processed by {\it executables}, whose core component is {\it logic}, which can be designed to integrate data obtained in desired ways. For example, with data sent from multiple nodes (namely individual oracles) to the DON, the logic of the executable can take the median of the data obtained to cancelled out the overly high or low data, which is likely to be malicious, and return the relatively accurate final result (the median) to users. The flowchart of Chainlink data processing is shown in Figure \ref{chainlink_label}.

\subsubsection{Incentivization Mechanism}

If data is requested from a DON node, the data providing nodes can charge certain gas fee in LINK token as their reward. Therefore, for data requesters, the more nodes used as data providers, the more expensive it is to aggregate the data and the higher the reliability of the data received. Furthermore, Chainlink oracle rates the performance of data providers to as a quality evaluation (namely the nodes of DON), which can be regarded as a reputation system with evaluation criteria including {\it Total Transactions}, {\it Total Link Earned}, {\it Response Ratio}, {\it All Time Average Response (Blocks)} and so on \cite{GG1}. The higher the reputation ranking of a node is, the more likely users request data from it, which therefore motivates data providing nodes to act honestly and efficiently.

However, the system merely provides users with quantified index of performances of the nodes and does not further filtering or restriction on data providing nodes, but only rely on the choice of user group. Therefore, although with reputation ranking among data providers, Chainlink is not regarded as a typical reputation-based data processing oracle but still an aggregating-based data processing oracle.

\subsubsection{Applications and Ecosystem}
Currently, Chainlink is one of the largest decentralized oracles with application in DeFi projects including lending (such as Aave and dYdX), synthetic assets (Synthetix), decentralized exchange (Pancakeswap), etc. as introduced in its official website. The market capitalization of LINK token reached more than 12 billion USD by the time of writing.

\subsection{Band Protocol Oracles\cite{HH1}}

\begin{figure}
\centerline{\includegraphics[width=20pc]{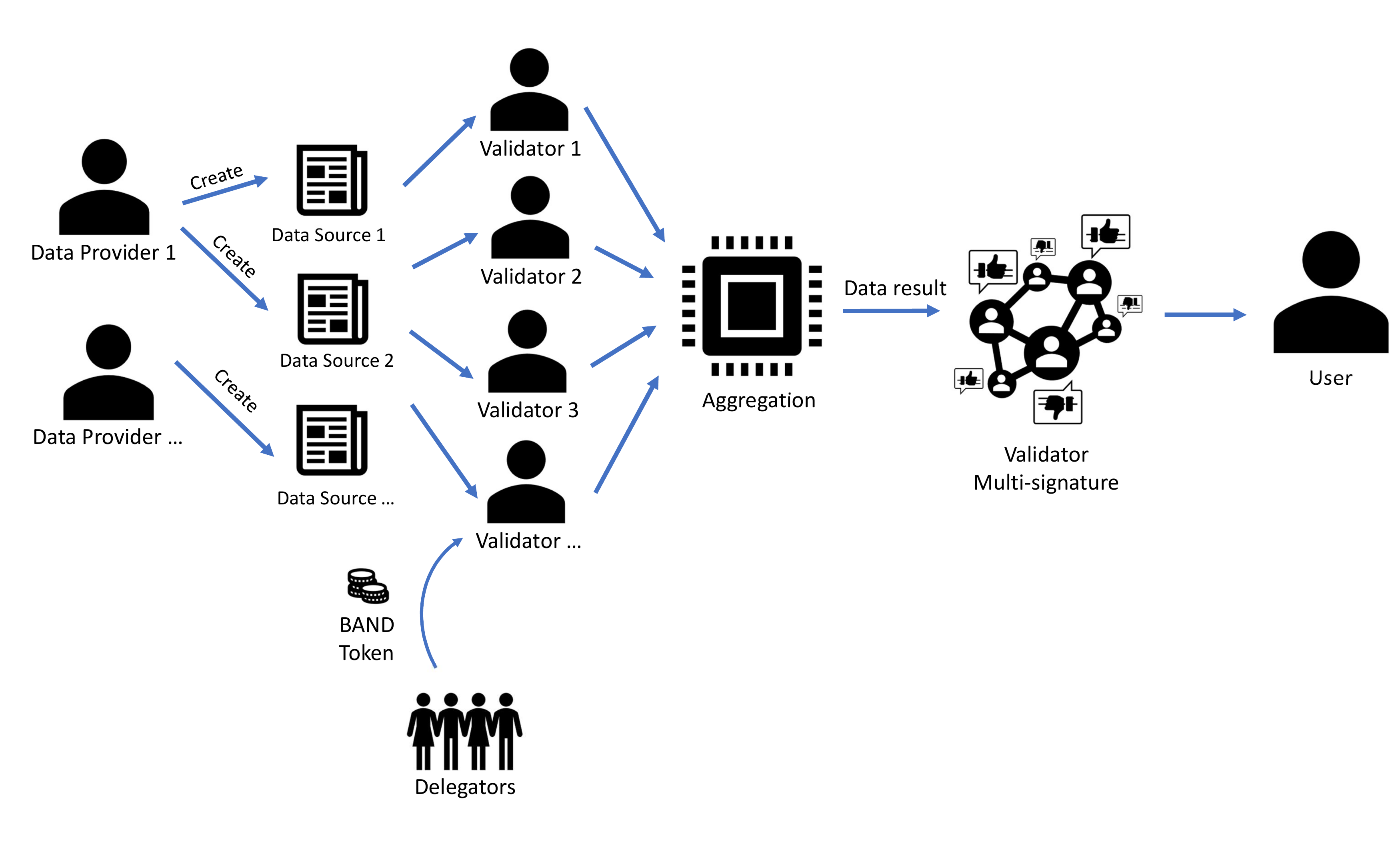}}
\caption{Flowchart of Band Protocol data processing}
\label{bandprotocol_label}
\end{figure}

Band protocol oracle is a staking-based decentralized oracle. It has certain similarity with Chainlink oracle on data aggregation procedure, but also difference mainly lying in the incentivization mechanism (namely the staking-based data processing).

\subsubsection{Design Criteria}
According to the whitepaper of Band protocol\cite{HH1}, the goals of Band protocol includes {\it speed and scalability} (being able to process large amount and high throughput data), {\it cross-chain compatibility} (achieving blockchain-agnostic system with zero-trust and high efficiency data verification) and {\it data flexibility} (building a permissionless and open system that is compatible with future changes and new data aggregation methods).

\subsubsection{System Architecture}
Users can participate in the network in 3 roles: data providers, validators, and delegators. Data provider can be any user that creates a data source (a data fetching script from pre-determined source) on BandChain. Validators mainly contribute by proposing and putting new blocks onto the chain and responding to data requests from oracle clients by obtaining data from data sources. Delegators are who do not commit to data on-chaining but stakes their tokens at the validators to receive commissions from them. With this delegation mechanism, BAND Protocol is a Delegation Proof-of-Stake system (DPoS).

Band protocol allows users to interact with the oracle through a Cosmos¡¯ {\it Inter-Blockchain-Communication} (IBC) protocol, therefore enabling other IBC-compatible blockchains for the oracle.

\subsubsection{Data Validation Process}
In a data feeding process, after the clients request for data, validators obtain data from data sources, and on-chain the data to the system for aggregation. The data from data sources is therefore aggregated by the system.

After a result of aggregation is obtained, validators verify the result on new blocks. Like most Cosmos blockchains, a new block is validated through multi-signature, and therefore new price data produced. The flowchart of Band Protocol is shown in Figure \ref{bandprotocol_label}.

\subsubsection{Incentivization Mechanism}
Band Protocol has its native token, BAND token with 7\% - 20\% inflation rate each year, enabling the users to participate in governance and to obtain reward fee for processing transactions.

In order to become validators, users need to stake their BAND at the oracle, and the probability of being selected as validators is proportional to the amount of BAND staked. A validator gets rewards in BAND for performing tasks including provisioning new blocks of the chain and processing transactions. Validators may be slashed (meaning their stake fined by the system) due to adversarial actions including participating too few block proposals and commits, double signing or unresponsiveness for data requests. For delegators, they are incentivized to stake BAND token to obtain rewards. For data providers, they are able to participate as {\it Premium Data Provider} to receive reward from creating data sources.

\subsubsection{Applications and Ecosystem}
According to the official website of Band Protocol, it has been partnered to DeFi projects including Terra, Mirror Protocol, etc.. It also supports various DeFi exchanges, including Binance, Uniswap, Coinbase, etc.. The native BAND token of the oracle has reached a market capitalization of 340 million USD by the time of writing.

\subsection{NEST Protocol \cite{II1}}

\begin{figure}
\includegraphics[width=20pc]{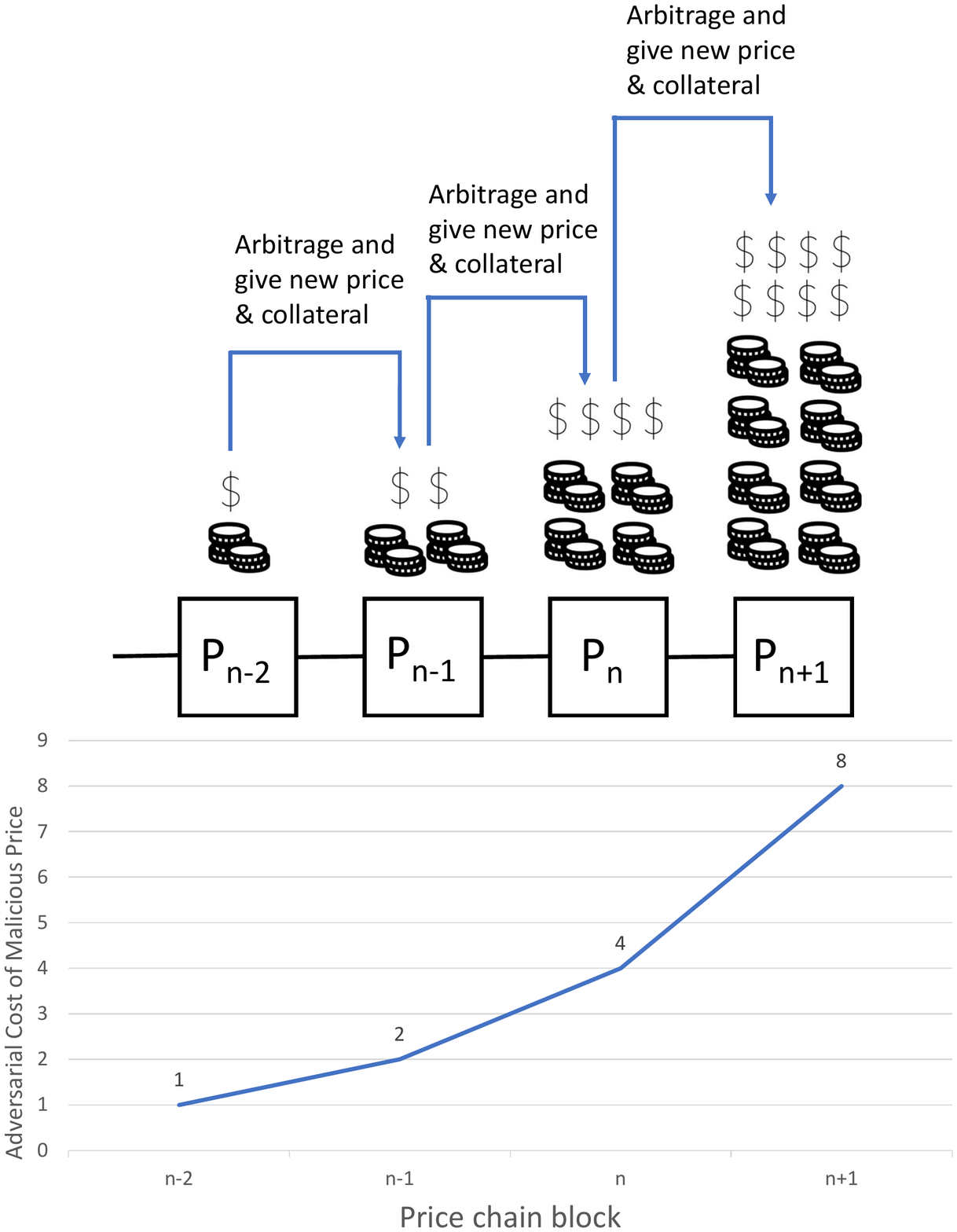}
\caption{Price chain and collateral of NEST Protocol }
\label{nest_label}
\end{figure}

NEST Protocol is a game-theory-based data processing oracle, which aims at economically disincentivizing users from malicious behavior. Compared to staking-based processing, its incentivization process only exists during the process of data providing and does not require pre-existing staking of participants in the system or slashing after data providing, and any user can join the process without prerequisites.

\subsubsection{Design Criteria}

NEST Protocol points out in its whitepaper \cite{II1} 5 key criteria to judge oracle data quality, which are accuracy, price sensitivity (following changing data on time), attack resistant (high attacking cost), direct verification (data verifiable by any third party) and distributed quotation system (no requirement on user side and users are free to join and leave without malicious impact to the system).

\subsubsection{System Architecture \& Data Validation Process}
NEST Protocol is a game-theory-based data processing oracle, with a core mechanism of {\it Quote Mining Process}, a decentralized incentive solution for price reporting.

The quote mining process of Oracle Quote mechanism enables users to report new prices between 2 tokens and correct unreasonable prices (as shown in Figure \ref{nest_label}). The process has following steps:

1. Any participant can pass a price among 2 assets to the quotation contract, meanwhile collateralizing certain amount of the 2 assets to the system at a proportion according to the price just reported.

2. After step 1., the system will observe the price for T0 time, which is 25 blocks (around 5 minutes). During this period, anyone is able to liquify a part or all the previous collateralized assets at the price if the price is considered to enable arbitrage compared to market price. This means unreasonable price will be arbitraged by later upcoming users. If during T0 period, all of the collaterals are liquified by other users, and total collateral left is zero, the price will be invalid; and vice versa if some transactions happen and part or none of the assets collateralized are liquified, the price is considered valid and the amount of collateral is remaining value left not liquified during T0 period.

3. The users who arbitrage at the price during T0 have to report a new price to the system, and leave the collateralized assets, which is required to be beta times of the amount he traded at the original price. Currently, beta is 2, which means the adversarial cost doubles at each new price and increases geometrically.

4. The new price is reported and new assets collateralized form a chain, namely the price chain.

\subsubsection{Incentivization Mechanism}

For a miner (namely who propose new blocks in the price chain), the cost of mining a price block on the chain is 1\% of the quoted ETH scale and the gas fee and as a reward NEST token will be issued to miners who successfully on-chains a price chain block. The total amount of NEST token release has an upper bound of 10 billion, which will be 400 release each block at the beginning and decrease to 80\% of the last period every 2.4 million blocks (around 1 year), and after the block reward is reduced to 40 NEST per block the release will no longer decrease.

\subsubsection{Applications and Ecosystem}
NEST Protocol currently has partnership with exchanges including CoFix, crypto.com, iNFT, etc. according to its official website. The market capitalization of its native token NEST has reached 25 million USD by the time of writing.

\subsection{DOS Network\cite{CC1}}

\begin{figure}
\centerline{\includegraphics[width=23pc]{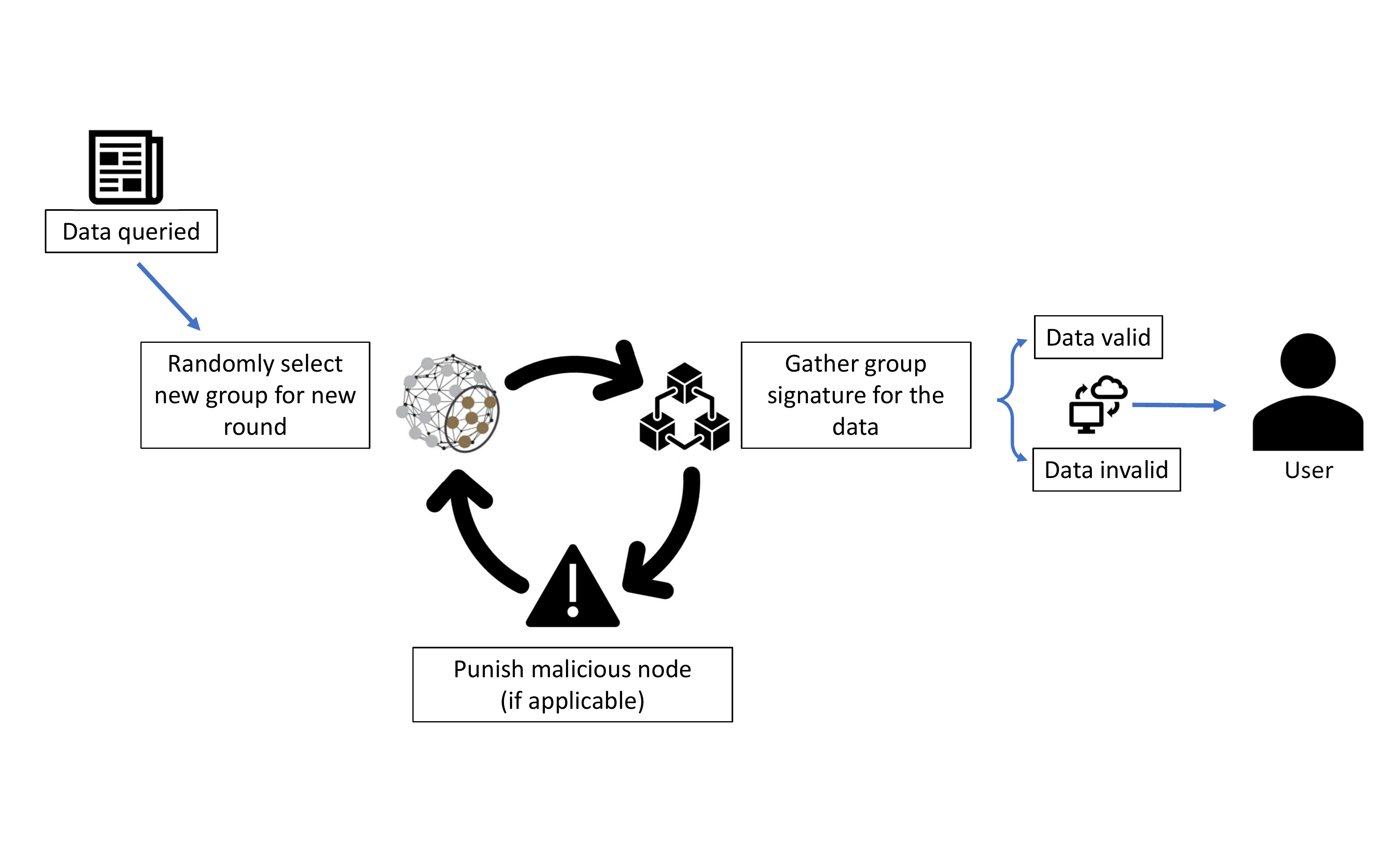}}
\caption{DOS Network working flowchart}
\label{dos_label}
\end{figure}

DOS Network is a reputation-based oracle, which decides the opportunity of participation of a user with a reputation score mechanism. Different from Nest Protocol mentioned above, reputation-based oracle does not only give a short-term penalty to malicious users (collateral liquified), but also long-term penalty of losing change of participation in the system.

\subsubsection{Design Criteria}
According the Whitepaper, DOS Network offers mainly 2 aspects of service, which are decentralized data feeding (by providing connection among on-chain and off-chain data) and decentralized verifiable computation (by providing computation power to blockchains). Based on these services, DOS Network enables chain-agnostic oracle and Dapp construction.

\subsubsection{System Architecture}
The Decentralized Oracle Service (DOS) Network is a reputation-based network based on layer-2 protocols, providing services including decentralized data feeding and decentralized verifiable computation oracles.

Its design mainly consists of 2 parts, the on-chain part, and the off-chain part of the oracle. The on-chain part of the oracle includes a proxy system and on-chain governance system (consisting of monitoring, registration, and payment systems). The proxy system is an interface for the user to interact with on-chain user contracts, and the governance system mainly provide services including recording the Quality of Service (QoS) of DOS nodes (namely monitoring), registering new joining DOS nodes and processing payment rewarded to the DOS Network nodes runners.

The off-chain part of its design mainly guarantees the validity of the data provided by the network nodes, which is based on 2 techniques, the Verifiable Random Function (VRF) and the threshold signature scheme, and together they support a Byzantine Fault Tolerance (BFT) data feeding system of DOS Network.

\subsubsection{Data Validation Process}

DOS nodes need to deposit certain security reserve as collaterals in the system, and during data feeding, they are randomly divided into groups by the VRF, and a group is randomly selected to perform computation or to execute a configured script such as responding to data requests. Within the group, nodes can reach an in-group BFT consensus if the number of non-adversarial nodes is above a threshold (i.e. ''t-out-of-n''). The malicious nodes, if not responding or providing invalid data, it will be excluded from future runs of the protocol, and their security reserve will be slashed.

Each group has a quality score as a measure of their Quality of Service (in the terms of criteria such as correctness and responsiveness), and for non-responding groups, their negative score will be incremented until their score is abnormal to be banished from the protocol. Furthermore, for a node of overly low QoS, it will excluded from the off-chain protocol and payment process. The flowchart of DOS Network is shown in Figure \ref{dos_label}.

Therefore, a reputation-based data validation process is constructed with this QoS system.

\subsubsection{Incentivization Mechanism}
DOS Network has its native token, DOS Tokens, used to incentivize participants of the system, including Dapp developers (for submitting development proposals), mining node runners and premium data providers (those who monetize data feeding in the system). Nodes on-chaining a new block obtain rewards, while users can also obtain rewards from staking itself. If the user conducts adversarial actions, the penalty is being excluded from future participation, incentivizing honest actions.

\subsubsection{Applications and Ecosystem}
DOS Network currently has strategic partners including mining pools (Huobi Pool), lending platform (ForTube), blockchain infrastructure (Meter), etc. according to its official website. The market capitalization of its native token DOS has reached 40 million USD by the time of writing.

\begin{table*}[t]
\centering
\caption{Comparisons among Current DeFi Oracles}
\begin{tabular*}{17.1cm}{p{2cm}p{2cm}p{4cm}p{4cm}p{2cm}}
\specialrule{.1em}{.05em}{.05em}
& Trust model & Trustworthiness Source & Data Provider Scalability & Native Token \\
\specialrule{.1em}{.05em}{.05em}
Chainlink & Aggregation & Non-adversarial majority & Adaptor \& Hardware requirement & LINK\\
\hline
Band Protocol& Staking \& Aggregation & Non-adversarial majority \& high adversarial cost & Data Provider: data source account and infrastructure setup; Validator: staking & BAND\\
\hline
Nest Protocol & Game-theory & High adversarial cost & Any user can participate with low barrier & NEST\\
\hline
DOS Network & Reputation & Non-adversarial majority \& high adversarial cost & Staking & DOS\\
\hline
Witnet & Reputation & Non-adversarial majority \& high adversarial cost & Any user can join as a witness & WIT\\
\specialrule{.1em}{.05em}{.05em}
\end{tabular*}
\label{real_example_table_label}
\end{table*}

\subsection{Witnet Oracle\cite{QQ1}}

Witnet Oracle is a reputation-based oracle. Compared to DOS Network, its difference lies in the reputation score system that the total reputation score of the system is constant, and the increase of reputation score of some user can only be sourced to the decrease of others.

\begin{figure}
\includegraphics[width=18pc]{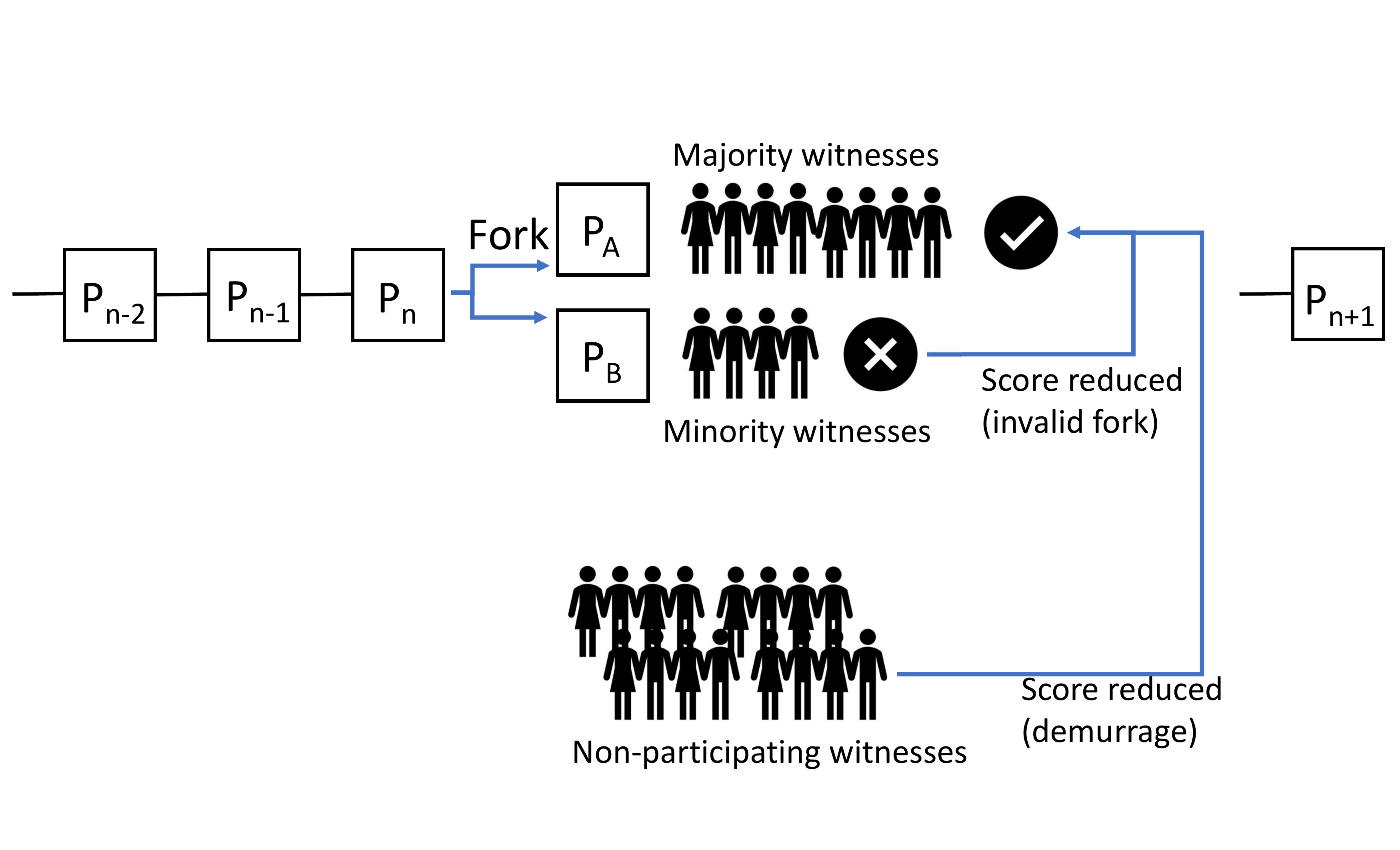}
\caption{Witnet oracle working flowchart}
\label{witnet_label}
\end{figure}

\subsubsection{Design Criteria}
Witnet Oracle is a reputation-based oracle giving data reporters reputation points based on their performance of data provision. The system was designed to conduct tasks of {\it Retrieve-Attest-Deliver}, to provide trustworthy data to clients.

\subsubsection{System Architecture}
In Witnet Oracle, users are divided into 3 types: clients, witnesses, and bridges. Clients are users who request certain data provision, witnesses work as data reporters, and bridges work as interconnection between the Witnet oracles and other public smart contract platforms. Witnet¡¯s data processing system is based on its {\it Decentralized Oracle Network} (DON).

In Witnet, prices are produced in the form of giving mining blocks on the price chain. If at a block there exist dispute among witnesses, the price chain will fork and proceed to the dispute solving by choosing the fork with majority reputation support as the valid result.

\subsubsection{Data Validation Process \& Incentivization Mechanism }
Witnet incentivizes users with reputation scores and its native token Wit. Miner publishing a new block can obtain block reward in the form of Wit token, which is inflationary to provide long-run incentive. Meanwhile, the likelihood of successfully mining a new block is related to the reputation score he gets. Namely the higher the reputation score he gets, the more likely he is to get tasks, where such task returns payoffs to the witnesses.

Each witness (public key) has a minimum score, and at checkpoint 0 (the genesis block) of the ledger, the minimum score is set to be 1. Considering that the public key is of 256 bits, there exist $2^{256}$ initial score to be distributed among the witness identities (public keys). In additional to the basic 1 point, another x*$2^{256}$ points are used for re-distribution among the witnesses. Namely a witness starts with a reputation score of x+1 \cite{AA2}. If a witness loses all of his re-distributable points, he still has the minimum 1 score which is never reduced from him. In such system, the gain of reputation score of a witness only comes from the reduction of the scores of other witnesses.

If different data is given by different witnesses, they will form forks on the price chain, and witnesses anchor their reputation to different forks of the chain. Eventually, the data fork receiving the support of majority witnesses will become valid and survive as the final result, while other forks became invalid. For a witness who support an invalid fork, his reputation score will be reduced. The reputation scores reduced will transfer from dishonest witnesses to honest witnesses, while remaining the total amount to be constant. The flowchart of Witnet oracle is shown in Figure \ref{witnet_label}.

In addition, for a witness that does not participate in the data validation process (namely not anchoring his score to any of the data outcomes), his score will still be reduced by a demurrage process, and the more scores he has the faster his score reduces due to irresponsiveness. Therefore, a witness is incentivized to join the price reporting process.

Furthermore, Witnet system gives a so-called ¡°Double-agent Incentive¡± to the witnesses, meaning in the system witnesses do not reveal which claim of data they support. This means that even though bribers may provide economic incentives to witnesses, the witnesses can still choose to behave honestly to the DON and obtain both the reward from the system and the bribers.

In this case, it is important to find the optimized balance between the dishonesty penalty and the demurrage. If the former is too low, the quality of data providing cannot be guaranteed due to low adversarial cost; if the latter is too low, the witnesses may tend to not participate the price reporting process to avoid the risk of supporting the wrong forks, causing the operation of the system inefficient. Therefore, a balance is required so that the witnesses are incentivized to report data to the client honestly in such mechanisms.

\subsubsection{Applications and Ecosystem}
According to its official website, Witnet¡¯s ecosystem mainly includes other related infrastructure developed by Witnet, including wallet (Sheikah wallet) and Witnet.network block explorer. Currently, the native token of Witnet, WIT, is not active in DeFi market.

\begin{figure*}
\centering
\includegraphics[width=\textwidth]{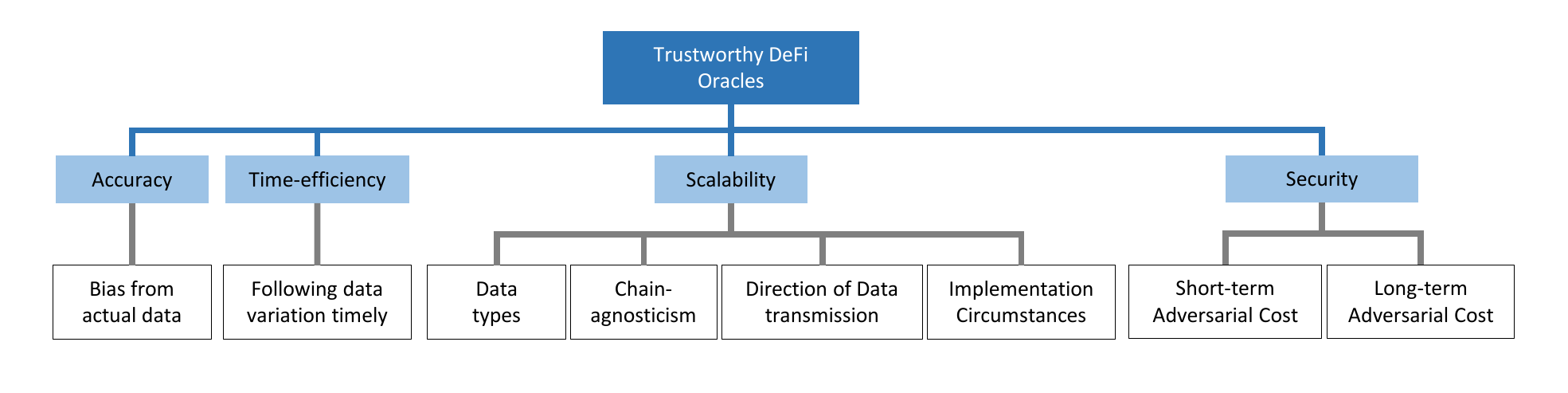}
\caption{Design Metrics for DeFi Oracles}
\label{future_criteria_label}
\end{figure*}

\subsection{Comparisons: Similarities and Differences }
In previous sections of this paper, decentralized oracle designs are categorized into 4 types, and essentially, their source of trustworthiness can be categorized into two types: (1) non-adversarial majority, and (2) benefit of adversarial acts smaller than benefit of honest acts.

Aggregation-based type mainly rely on (1) and barely relying on (2), since it merely aggregates the data received in a node-agnostic and data-quality-agnostic way with pre-determined algorithms. Staking-based type also mainly relies on (1), but compared to aggregation-based type it rely more on (2) since it provide economic incentive to voters by staking, rewarding and slashing, and the higher staking is the less likely the user is to vote for a malicious price. Game-theory based type mainly rely on (2) rather than (1), and since it barely rely on the proportion between adversarial and honest users, it can work in environments with more malicious users. Reputation-based type relies on both (1) and (2), since reputation incentives the users with future participation opportunities and it also decide the reputation of a user with its performance compared to the majority users. For various oracle designs, their suitable circumstances of applications depend on the feature of their source of trustworthiness. Therefore, judging the source of trustworthiness from criteria mentioned in the session above, Chainlink mainly relies on (1), NEST Protocol mainly relies on (2), and Band Protocol, DOS Network and Witnet rely on both (1) and (2) as source of trustworthiness.

With respect to the scalability of the data processing nodes, current active oracles have different characteristics. For Chainlink, users need to use adaptors of Chainlink and fulfill hardware requirement to become a data feeding node. For Band Protocol, to become a data provider, a Data Provider¡¯s Account and a ¡°gateway server¡± is required to be set up, and to become a validator, staking is required on the system. For DOS Network, users are also required to stake to become data validator, while for NEST and Witnet, any user is able to join the data validation process as long as they are in the network. The comparison above is summarized to Table \ref{real_example_table_label}.

Furthermore, it cannot be ignored that new data validation mechanism may show up in the future. For example, Uniswap, as a decentralized exchange, has the potential of being converted into an oracle with its current infrastructure, according to Vitalik Buterin\cite{DD2} since Uniswap has one advantage of large market capitalization, which increases the adversarial cost of attackers.

Although with current data validation mechanism, current oracles are not solving the demand to oracle problem perfectly and is with some shortcomings. For example, inaccuracy of price data feeding causes huge loss for DeFi platforms and users; data feeding may not be on time; data type requested by users may not be reachable by the oracle or the platform due to technical issue; DeFi attacks by manipulating oracles are frequently heard, etc.. The problems DeFi oracles are facing gives us insight on the potential development of them in the future in the next section.

\section{Future of DeFi Oracles}

With past and present DeFi oracles analyzed and compared, in this section, we give out our view on the future of trustworthy DeFi oracles, including the metrics of DeFi oracles and the trustworthiness mechanism for data feeding and data providers.

\subsection{Metrics for Building Trustworthy DeFi Oracles}

Current DeFi oracles have disadvantages in the application to the industry, as \cite{JJ2} pointed out, oracle problems may encounter malfunctions, biased data and lack of timeliness of data feeding, which reflets some possible future direction of improvement. In this paper, we propose some possible improvement metrics or criteria of oracles, which include accuracy, time-efficiency, scalability, and adversarial cost (risks). (as concluded in Figure \ref{future_criteria_label})

\subsubsection{Accuracy}
Accuracy means that the variance between data fed by the oracle and actual data is held in an acceptable threshold. The variance may be caused by the fluctuation of price or the difference among difference data sources and trading price of different markets. In the highly fluctuating DeFi market, small proportion of price change may cause relatively large fluctuation of DeFi asset value.

Currently, DeFi oracles have limited accuracy. For example, Chainlink updates the price when a deviation threshold of the price is surpassed. Within the deviation threshold, the price with certain inaccuracy may still cause loss of the users. Therefore, oracles have to take the capability of providing accurate enough data for the data requesters as one important criterion.

\subsubsection{Time-efficiency}
Time-efficiency measures the timeliness of data fed to users when receiving data request. With changing data source, whether oracles can update the data on-time is an important factor of whether the oracle provide qualified data feeding in fast fluctuating DeFi market.

In the future the importance of balance between accuracy and time-efficiency in the future rises in oracles since the demand of which is rising, while higher accuracy means more time spent on data validation, and less data processed in limited time period and in contrast time-efficiency may requires less data validation time.

\subsubsection{Scalability}
Scalability means that oracles are able to integrate and adapt to more data types and chains and with more flexible functionality. With more scalability, users may receive and transmit data in more circumstances and occasions, expanding the market and user group of DeFi.

Currently, we can see such trend on 3rd generation blockchains such as Polkadot chain, which enables chain-agnostic data transmission\cite{OO1}. Outbound oracles (also known as reverse oracles) report on-chain data to off-chain algorithms, in contrast to traditional oracles. Parsiq\cite{JJ1} is nowadays one of the example of application of outbound oracles, providing services including on-chain wallet supervision and management. In addition, DeFi oracles are being applied to more circumstances, with more types of DeFi projects and ecosystems, and integration with traditional centralized finance.

\subsubsection{Security}
For an oracle, degree of security is reflected by the cost of adversarial users conducting malicious actions and attacks, which is an important factor to improve the reliability of oracles since it is directly related to the probability of attacks.

Adversarial costs can be divided into short-term and long-term ones. Although now traditional oracles including Chainlink and Band Protocol are dominating the market with short-term costs (including cost of 51\% attack and stake slashing), we can see new designs coming up with more long-term way to raise adversarial costs. For example, reputation-based oracles are especially potential in this field since it risks the long-term future chance of participation of the malicious users on the reputation system.

\subsection{Proposed Trust Architecture for Future Trustworthy DeFi Oracles}

\begin{figure*}
\centering
\centerline{\includegraphics[width=\textwidth]{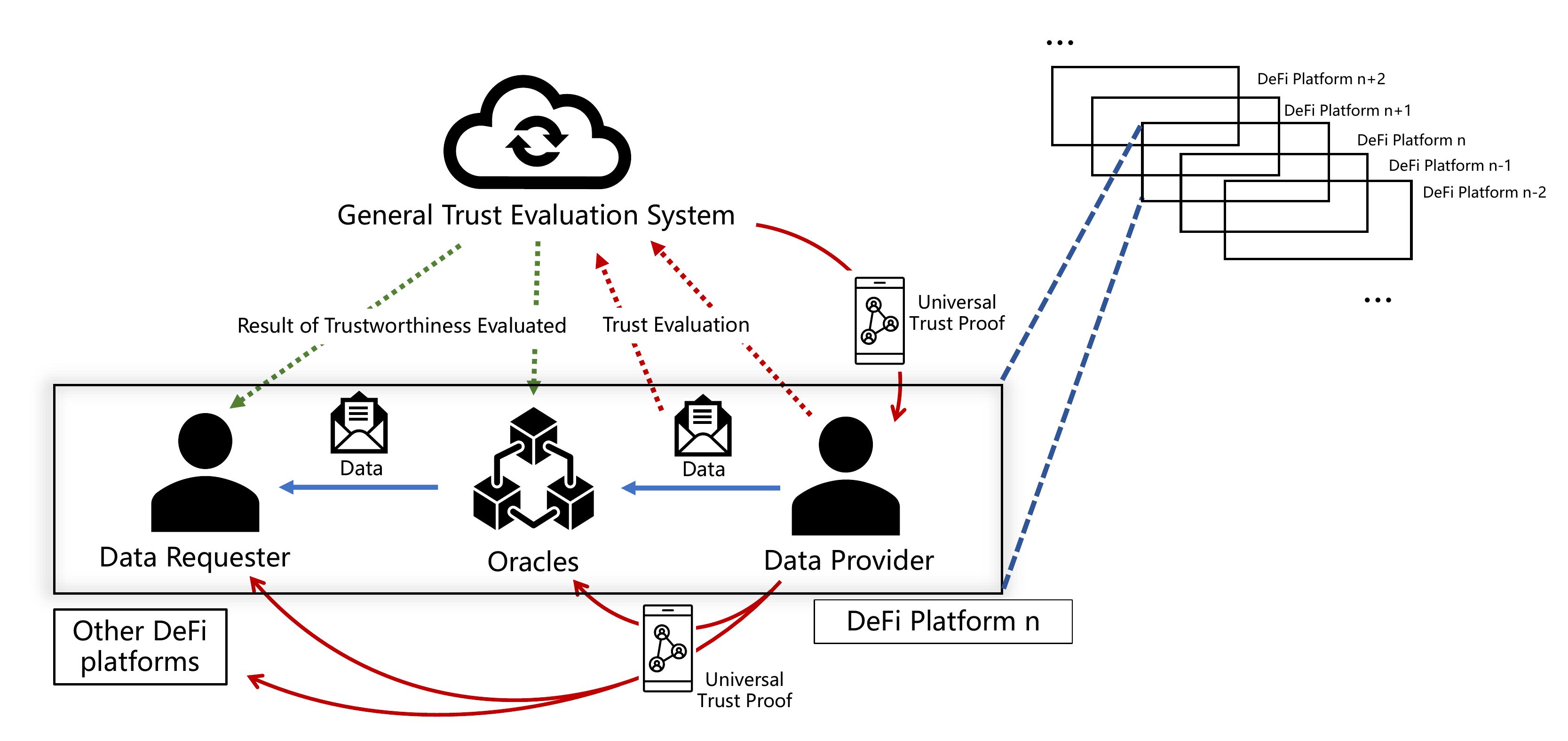}}
\caption{Blueprint for Future DeFi Oracles}
\label{blueprint_label}
\end{figure*}

With potential improvement on the metrics of DeFi oracles mentioned above, oracles will not only evaluate the trustworthiness more efficiently, but also make trust evaluation more universal and general among DeFi industry. It is possible for a user possessing certain universal reputation proof among difference blockchains and DeFi platforms, supported by a general trust evaluation system. (flowchart for demonstration as shown in Figure \ref{blueprint_label})

\subsubsection{On the Trustworthiness of Data Feeding}

With metrics mentioned above for DeFi oracles, data feeding is challenged with higher standards. Currently, DeFi oracles mainly relies on users¡¯ group behavior to decide the trustworthiness of data, which may not be the optimal approach with respect to time-efficiency and operational cost since extra economic incentive are required for user group to act honestly rather than adversarial.

\paragraph{Automated Trust Modelling}

Automated algorithm for trustworthiness is one of the possible future development directions, where trust modelling is the key factor for trustworthiness evaluation. In \cite{EE2}, Lim et al categorized trust models into 4 basic categories, including basic models, such as weighted average, relational measurement, graph methods, Bayesian methods and machine learning method, etc.. If applied to DeFi, such trust modelling is implementable and automatable by algorithms (namely smart contracts in DeFi), largely automating the trust evaluation on the data, rather than evaluating the reputation of user through the group behavior of majority users.

\paragraph{Machine Learning in Data Validation}

Data validation is the process of determining the trustworthiness of data and one of the promising approaches of automated trust modelling is machine learning trained by historical data and peer projects, etc.. Compared to the approach supported by user group behavior, machine learning approach is able to process much larger amount of data than user committees do. Furthermore, machine learning trust model is relatively more scalable and migratable to other systems as long as new data set or environment is provided, while in contrast human users conventionally tend to process certain data type or certain platform. With respect to user cost, a machine-based trust model does not require economic incentive to make honest decision and avoid adversarial ones, which lowers the system operating cost and makes the system less corruptible.

Furthermore, the data used for machine learning trust model is highly accessible since DeFi users¡¯ (namely the on-chain address) trading records are publicly traceable, obtaining information such as amount of the transaction, blockchain address of sender and receiver, platform of transaction, etc.. Such information can be useful as input data for a classification on whether the transaction was an attack or whether the user is trustworthiness in certain circumstances, which in contrast is not feasible to implement by human effort on each data provider due to the effort required.

\subsubsection{On Trustworthiness of Data Providers}

Beside the trustworthiness of data itself mentioned in last section, the trust on data providers is also of importance. Reputation-based oracle has certain advantage over other oracles in this aspect. It is a more direct and essential solution to the problem and will be further discussed in this session.

\paragraph{Reputation-based Oracles: Limitation and Advantage}

Currently, reputation-based oracles have certain limitations. As section E-1 of \cite{MM1} pointed out, reputation-based data validation system may be less cost-effective. Such disadvantages appear in human-based oracles described in \cite{MM1}, where users experience opportunity cost of giving up benefit obtained from adversarial behavior and need extra incentive in order to overcome the opportunity cost, which makes the system run less cost-effectively.

Reputation-based oracles¡¯ advantages includes but are not limited to: i) High barrier for adversarial user participating the system. Reputation system usually limits the chance for low-reputation users to participate in the data feeding process of the oracle system and requires much more effort to gain more reputation than losing it, which is also known as a property of trust, {\it easier to lose than to gain}. ii) Reducing malicious data entering the system effectively. By filtering out data provided from low reputation users, malicious data cannot enter the system. For oracles based on other data validation mechanisms, adversarial users can still participate in data processing but only with their impact restricted or disincentivized, leaving higher potential threats to the system. iii) Introducing reputation and trust into the system fits future trends of development. With more integration with traditional finance and physical world, a more stable investment and trading environment is demanded, especially with concepts introduced from CeFi, including KYC (Know Your Costumer), AML (Anti-Money Laundering), CFT (Combatting Financial Terrorism), etc.. Reputation and trust system is able to fill the gap between traditional centralized and on-chain world.

\paragraph{Reduction of Human Effort in Reputation Evaluation}

As mentioned in previous sections of trust on data feeding, one direction of future development is conducting user reputation evaluation with less human factors and automation. Currently, the trust on data by reputation-based oracles is based-on its provider¡¯s reputation, which is evaluated by the general behavior of user group, for example, Witnet choose the data fork with the majority witnesses¡¯ support as the one with valid data and DOS gives negative QoS score if the data provided deviates from majority.

Such process has limited effectiveness. On one hand, it always requires some assumption on the general user group, for example, that the proportion of adversarial users does not exceed certain threshold (50\% in the case of Witnet) so that the system can always return valid answer. On the other hand, the quality of data does not necessarily depend only on the reputation of its provider. Therefore, it is reasonable to reduce the reliance on human behavior and introduce more factors that can be observed by automated algorithms to determine the trustworthiness of data, for example, the time distance of the data being published from current time, the response speed and rate of the user, etc..

Furthermore, reputation evaluation decides whether certain user is trustworthy or not, which is essentially a classification process, and the adoption of machine learning into reputation system is worth attention. With determining factors of reputation known, the algorithms to compute reputation from those factors still requires careful consideration of oracle developers. For example, when given data of trust factors A, B and C relevant to evaluate the reputation of a user, different weights (x, y and z) can be given to A, B and C to form a calculation: xA + yB + zC. Due to the complexity and fast-changing property of DeFi environment, optimizing the weights of the calculation with human effort may raise the operational cost and cause potential inaccuracy or time-inefficiency to the reputation system. Machine learning can largely save the effort and provide outcomes in a more time-efficient and accurate way as demonstrate in previous sections.

\subsubsection{General Trust Evaluation System}
The trend can be predicted that, in the future trust evaluation has the potential of generalization not limited to one specific chain or DeFi project, due to the demand of transactions and user migration among different chains and platforms. Such evaluation system may be supported by specialized oracles and smart contracts obtaining related data on the trustworthiness of users and conducting trust evaluation through certain algorithm and trust model.

Trustworthiness can be evaluated during the transmission from data providers (including APIs, human, smart contracts, etc.) to requesters (smart contracts, human users, etc.), by conducting trust evaluation on both the data itself and the provider of it.

\subsubsection{Universal Reputation Proof}
With a generalized trust evaluation system mentioned above, it is likely that users or other data sources may possess certain universal proof of their reputation to help data receivers validate the data fed, especially for DeFi projects without trust evaluation system of their own. Such universal proof is able to prevent users from conduct malicious actions as a ¡°new user¡± in a different DeFi platform without receiving penalty. For example, it is possible to represent the reputation of a user with an NFT (non-fungible token), which is unique as a single proof among the users, enabling reputation evaluation universally.

It is possible that such reputation proof is not limited to DeFi world. For a DeFi project that is related to real world assets, for example synthetic assets of households, reputation proof may be linked to the real-world identity with zero-knowledge publicized to the network. Such feature may contribute to the connectivity between DeFi and traditional centralized finance.

\section{CONCLUSION}

In recent years, decentralized finance (DeFi) has appeared as a rapid developing field, where oracles provided viable solutions and promising applications. In this paper, the applications of oracles in DeFi has been introduced, including DeFi lending, synthetic assets and insurance. The past development of DeFi oracles has been introduced by categorizing them into the aggregation-based, staking-based, game-theory based and reputation-based based on their data processing features. Furthermore, five current active oracles have been introduced with respect to their system architecture, data validation process and incentive mechanisms, and a detailed comparison has been conducted among them according to the trustworthiness of the data and its sources as well as their overall trust models. Lastly, metrics and possible future techniques have been proposed, including the application of automation and machine learning, and a potential overall trust architecture has been given.


\EOD
\end{document}